\documentclass[%
 reprint,
superscriptaddress,
 amsmath,amssymb,
 aps,
]{revtex4-1}

\usepackage{graphicx}% Include figure files
\usepackage{dcolumn}% Align table columns on decimal point
\usepackage{bm}% bold math

\usepackage{graphicx}
\usepackage{MnSymbol} 
\usepackage{braket}
\usepackage{multirow}
\usepackage{hhline}
\begin{document}

\preprint{APS/123-QED}

\title{Coherent coupled qubits for quantum annealing}

\author{Steven J. Weber} 
\email{steven.weber@ll.mit.edu}
\affiliation{MIT Lincoln Laboratory, 244 Wood Street, Lexington, MA 02420, USA}
\author{Gabriel O. Samach}
\affiliation{MIT Lincoln Laboratory, 244 Wood Street, Lexington, MA 02420, USA}
\author{David Hover}
\affiliation{MIT Lincoln Laboratory, 244 Wood Street, Lexington, MA 02420, USA}
\author{Simon Gustavsson}
\affiliation{Research Laboratory of Electronics, Massachusetts Institute of Technology, Cambridge, MA 02139, USA}
\author{David K. Kim}
\affiliation{MIT Lincoln Laboratory, 244 Wood Street, Lexington, MA 02420, USA}
\author{Alexander Melville}
\affiliation{MIT Lincoln Laboratory, 244 Wood Street, Lexington, MA 02420, USA}
\author{Danna Rosenberg}
\affiliation{MIT Lincoln Laboratory, 244 Wood Street, Lexington, MA 02420, USA}
\author{Adam P. Sears}
\affiliation{MIT Lincoln Laboratory, 244 Wood Street, Lexington, MA 02420, USA}
\author{Fei Yan}
\affiliation{Research Laboratory of Electronics, Massachusetts Institute of Technology, Cambridge, MA 02139, USA}
\author{Jonilyn L. Yoder}
\affiliation{MIT Lincoln Laboratory, 244 Wood Street, Lexington, MA 02420, USA}
\author{William D. Oliver}
\email{oliver@ll.mit.edu}
\affiliation{MIT Lincoln Laboratory, 244 Wood Street, Lexington, MA 02420, USA}
\affiliation{Research Laboratory of Electronics, Massachusetts Institute of Technology, Cambridge, MA 02139, USA}
\affiliation{Department of Physics, Massachusetts Institute of Technology, Cambridge, MA 02139, USA}
\author{Andrew J. Kerman}
\email{ajkerman@ll.mit.edu}
\affiliation{MIT Lincoln Laboratory, 244 Wood Street, Lexington, MA 02420, USA}

\begin{abstract}

Quantum annealing is an optimization technique which potentially leverages quantum tunneling to enhance computational performance.  Existing quantum annealers use superconducting flux qubits with short coherence times, limited primarily by the use of large persistent currents $I_\mathrm{p}$.  Here, we examine an alternative approach, using qubits with smaller $I_\mathrm{p}$ and longer coherence times.  We demonstrate tunable coupling, a basic building block for quantum annealing, between two flux qubits with small ($\sim 50~\mathrm{nA}$) persistent currents. Furthermore, we characterize qubit coherence as a function of coupler setting and investigate the effect of flux noise in the coupler loop on qubit coherence.  Our results provide insight into the available design space for next-generation quantum annealers with improved coherence.  

\end{abstract}

\maketitle

\section{Introduction}

Quantum annealing~\cite{finn94,kado98,broo99,farh01} is a heuristic technique for finding the low energy configurations of complicated Ising models.  It has received considerable interest as a potential new computing paradigm for solving classical optimization problems~\cite{luca14}, which are important for a wide range of applications in science and industry.  Existing quantum annealers, despite rapid progress in system size and intensive efforts to benchmark performance, have yet to demonstrate improved scaling over classical methods~\cite{ronn14,boix14,lant14,alba15,katz15,isak15,boix16,denc16,mand16}.  While continued efforts to scale and improve existing quantum annealing architectures will provide a clearer picture of their potential capabilities, it is also worthwhile to consider their limitations and explore alternative approaches which may be more amenable to quantum-enhanced performance.

Commercial quantum annealers, developed by D-Wave Systems, are based on niobium flux qubits with relatively short coherence times and are designed to implement stoquastic Hamiltonians~\cite{brav08} with pairwise Ising couplings limited to a ``Chimera" connectivity graph~\cite{harr10,buny14}.  Experience with the D-Wave platform suggests that it could benefit from higher connectivity, increased precision in setting parameters, and greater control over the annealing schedule. In addition, increased qubit coherence, non-stoquastic Hamiltonians, and multi-qubit interactions~\cite{kafr16} may also be instrumental in achieving quantum-enhanced performance.  In this work, we focus on the challenge of improving the coherence of coupled qubits in a quantum annealer.

Superconducting flux qubits~\cite{mooi99,orla99} are well-suited to quantum annealing, because they can be used to approximately realize the transverse Ising model Hamiltonian $\hat{H}_I=\tfrac{\hbar}{2}\sum_i (\epsilon_i\hat{\sigma}^{\mathrm{z}}_i+\Delta_i\hat{\sigma}^{\mathrm{x}}_i)+\sum_{i<j} \hbar J_{ij}\hat{\sigma}^{\mathrm{z}}_i\hat{\sigma}^{\mathrm{z}}_j$, where $\hbar\epsilon_i$ and $\hbar\Delta_i$ play the roles of the Zeeman energies due to the $z$ and $x$ components of the local field seen by the $i^\textrm{th}$ spin, and $J_{ij}$ is the Ising interaction between spins $i$ and $j$. Here, the two eigenstates of the Pauli operator $\hat{\sigma}^{z}_i$ correspond to ``persistent current" states of qubit $i$, which can be viewed as clockwise and counter-clockwise currents of magnitude $I_\mathrm{p}$ circulating around the qubit loop. For a quantum annealing device based on the Transverse Ising model, the parameters $\epsilon_i$ and $J_{ij}$ are used to encode a classical problem, while nonzero $\Delta_i$ are the source of the quantum fluctuations which drive the annealing process. In general, all of these parameters must be tunable. In a flux qubit-based implementation, a coupler mediates an interaction between qubits $i$ and $j$, and the parameters $\epsilon_i$, $\Delta_i$, and $J_{ij}$ are tunable via local magnetic flux biases.

The coupler elements~\cite{mooi99,plou04,van05,nisk06,van07,zako07,harr07,ashh08, harr09,kafr16,allm10, allm14, chen14} are themselves also flux qubits, though operated in a regime where they can be described as a simple flux-tunable effective inductance $L_{\text{eff}}$. In this language, the coupling energy between two qubits, each with persistent current $I_{\mathrm{p}}$ and mutual inductance $M$ with the coupler, is given by $J = I_{\mathrm{p}}^2 M^2/L_{\mathrm{eff}}$.  The quantity $1/L_{\mathrm{eff}}$ is also referred to as the coupler susceptibility~\cite{harr09}.

The most significant design parameter to affect the coherence of a quantum annealer is the choice of $I_{\mathrm{p}}$. Flux qubits with large persistent current have a naturally strong coupling, as $J\propto I_\mathrm{p}^2$, but their coherence times are severely limited by flux noise: their sensitivity to flux noise is proportional to $I_{\mathrm{p}}$ and can limit both the energy relaxation time and the dephasing time~\cite{yan16,quin16}, which for $1/f$ flux noise roughly scale as $1/I_{\mathrm{p}}^2$ and $1/I_{\mathrm{p}}$, respectively.  In the D-Wave system, the qubits are designed with large persistent currents $I_{\mathrm{p}}\sim 3$ $\mu$A~\cite{harr10} in order to achieve large coupling strength with modest values of coupler susceptibility and $M$. In contrast, we recently demonstrated that it is possible to produce robust, long-lived flux qubits with small persistent currents $I_{\mathrm{p}}\sim 50$ nA~\cite{yan16}. In order to realize strong coupling between qubits with small $I_{\mathrm{p}}$, it is necessary to compensate by increasing either $M$ or the coupler susceptibility.  Although this approach increases the qubit's sensitivity to flux noise in the coupler loop and requires more precise control over the coupler flux bias, it nevertheless allows for a significant improvement in qubit coherence.

In this work, we demonstrate tunable coupling between qubits with persistent currents reduced by nearly two orders of magnitude compared to existing quantum annealers.  While coupled flux qubits with low persistent currents have been previously demonstrated \cite{van07}, no work to date has investigated the implications that this design choice has on qubit coherence for quantum annealing.  We present, for the first time, a systematic study of the coherence of coupled flux qubits in the context of quantum annealing.  In particular, we investigate the effect of flux noise in the coupler loop on qubit coherence.  Our results are in good agreement with simulations based on the full Hamiltonian of the coupled qubit system, as well as a semi-classical model. This work serves as a proof-of-principle and provides a framework for evaluating coherence in future quantum annealing architectures.   

\section{Experimental setup}

\begin{figure}
\begin{center}
\includegraphics[scale=1]{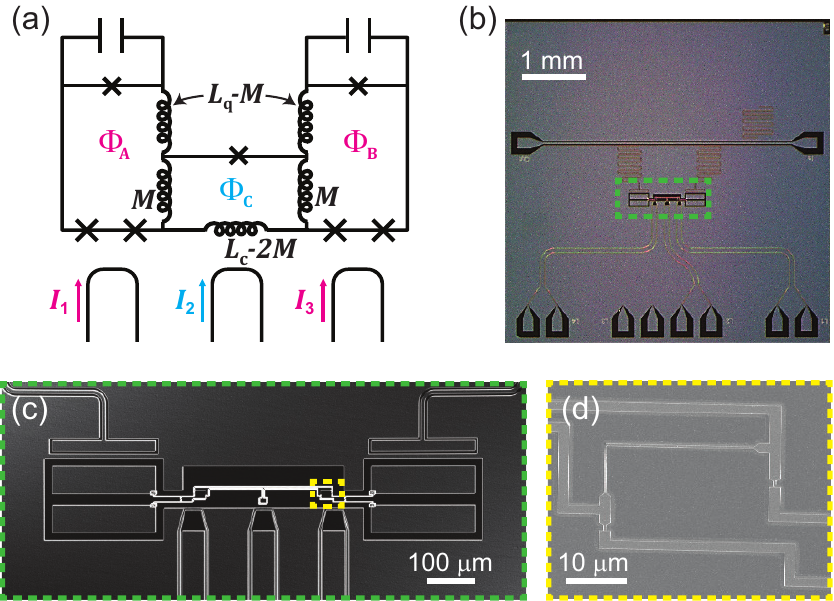}
\caption{\textbf{Coupled qubit geometry.} (a) Device schematic.  Qubit A (left loop) and Qubit B (right loop) are capacitively shunted three-junction flux qubits, coupled through a shared inductance with an rf-SQUID coupler (center loop). On-chip bias currents $I_1$, $I_2$, and $I_3$ control the external fluxes $\Phi_\mathrm{A}$, $\Phi_\mathrm{C}$, and $\Phi_\mathrm{B}$ through the qubit and coupler loops. (b) Optical micrograph of the aluminum (light grey) device on a silicon (dark grey) substrate. (c) Optical image showing the qubits, coupler, and flux bias lines. (d) SEM image of the galvanic connection between Qubit B (lower-right) and the coupler (upper-left).}
\label{fig:fig1}
\end{center}
\end{figure}

A circuit diagram of our coupled qubit device is shown in Figure \ref{fig:fig1}a.  Two capacitively shunted three-junction flux qubits, Qubit A and Qubit B, are each galvanically coupled to an rf-SQUID coupler via a shared inductance of $M = 34$ pH, as shown in Figures \ref{fig:fig1}b-d.  The devices are controlled by the externally applied magnetic fluxes $\Phi_\mathrm{A}$, $\Phi_\mathrm{B}$, and $\Phi_\mathrm{C}$.  For simplicity, our experiments use qubits with a single superconducting loop, instead of the multi-loop qubits that are required for independent $\Delta$ tunability.  We characterize the qubits using standard dispersive measurements~\cite{blai04}, with each qubit coupled to a separate readout resonator which is probed through a shared transmission line.

The transition frequency between the coupler ground- and first-excited state $\omega_\mathrm{C}^\mathrm{01}/2\pi$ was designed to be $\sim 20$ GHz, which is significantly larger than the qubit transition frequencies at $\sim 5$ GHz. Therefore, the coupled qubit system can be described by the approximate low-energy Hamiltonian $H \approx H_\mathrm{q}^{(\mathrm{A})} + H_\mathrm{q}^{(\mathrm{B})} + H_\mathrm{int}$~\cite{harr07}, where

\begin{align}
H_\mathrm{q}^{(i)} = \frac{\hbar}{2}[\epsilon_i(\Phi_\mathrm{A,B,C})\hat{\sigma}_\mathrm{z}^{(i)}+\Delta_i(\Phi_\mathrm{A,B,C})\hat{\sigma}_\mathrm{x}^{(i)}], \\ 
H_{\mathrm{int}} = \hbar J(\Phi_\mathrm{A,B,C})\sigma_\mathrm{z}^{(\mathrm{A})}\sigma_z^{(\mathrm{B})}.
\end{align}

\noindent The effective parameters $\epsilon_i$, $\Delta_i$, and $J$ are not only determined by the circuit parameters of the individual qubits and coupler, but also by their couplings, and can depend on all three flux biases. For each qubit, the degeneracy point is defined as the bias where $\epsilon_i = 0$. A table of device parameters can be found in Appendix A.  

\begin{figure}
\begin{center}
\includegraphics[scale=1]{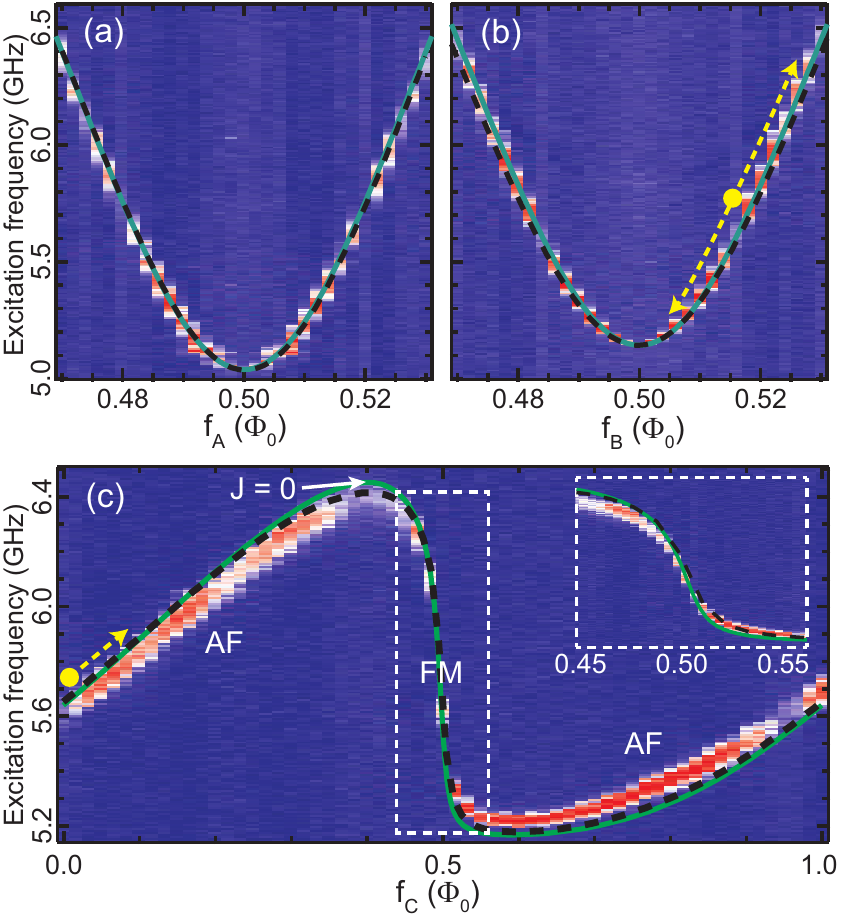}
\caption{\textbf{Qubit spectroscopy}. Dashed black traces: semi-classical model. Solid green traces: simulations of the full circuit Hamiltonian. (a) Spectroscopy of Qubit A vs $f_A$, with $f_B = f_C = 0$. (b) Spectroscopy of Qubit B vs $f_\mathrm{B}$ with $f_\mathrm{A} = f_\mathrm{C} = 0$. The yellow dashed line represents the starting point and range of qubit frequencies in the following panel. (c) Spectroscopy of Qubit B vs $f_\mathrm{C}$ for $f_\mathrm{A} = 0$ and $f_\mathrm{B} = 0.516$. The regions of anti-ferromagnetic (AF), ferromagnetic (FM), and zero coupling are indicated.  The inset shows detailed data for the FM region. }
\label{fig:fig2}
\end{center}
\end{figure}

The qubits were designed with shunt capacitance $C_{\mathrm{sh}} = 50$ fF, loop inductance $L_\mathrm{q} = 110$ pH, and $I_\mathrm{p}= 45$ nA.  All device components were patterned from a high-quality evaporated aluminum film on a high-resistivity silicon wafer, except for the superconducting loops and Josephson junctions, which were deposited using double-angle evaporation of aluminum~\cite{yan16}.   Spectroscopy plots showing the energy difference between the ground and first excited state for Qubit A and Qubit B are shown in Figures \ref{fig:fig2}a,b as a function of the reduced flux $f_\mathrm{i} \equiv \Phi_\mathrm{i}/\Phi_0$ in the qubit loop, with the coupler biased at $f_\mathrm{C} = 0$.  At this coupler bias, $\Delta_\mathrm{A}/2\pi = 5.042$ GHz and $\Delta_\mathrm{B}/2\pi = 5.145$ GHz.  

Figure \ref{fig:fig2}c shows how the transition frequency of Qubit B depends on the coupler bias.  This dependence originates from the circulating current in the coupler loop $\langle I^{\mathrm{C}}\rangle$, which couples to the qubit through the shared inductance $M$.  Thus, the coupler induces an offset flux in the qubit loop, which shifts the effective qubit bias as indicated by the dashed line in Figure 2b. Treating the interaction classically, the offset flux is given by $\delta f_\mathrm{B} = M \langle I^{\mathrm{C}}\rangle/\Phi_0$. Assuming that the coupler remains in its ground state, $\langle I^{\mathrm{C}}\rangle$ and $L_\text{eff}$ are related to the coupler ground state energy $E_0^{(\mathrm{C})}$ as

\begin{equation}
\langle I^{\mathrm{C}}\rangle \equiv \frac{\partial E_0^{(\mathrm{C})}}{\partial\Phi_\mathrm{C}} ;
\frac{1}{L_\mathrm{eff}} \equiv \frac{\partial \langle I^{\mathrm{C}}\rangle}{\partial\Phi_\mathrm{C}} = \frac{\partial^2 E_0^{(\mathrm{C})}}{\partial\Phi_\mathrm{C}^2}.
\end{equation}

By fitting our results to theory we extract the rf-SQUID coupler loop inductance $L_\mathrm{C} = 470$ pH and junction critical current $I_\mathrm{C}^0 = 730$ nA, giving $\beta \equiv 2\pi L_\mathrm{C} I_\mathrm{C}^0/\Phi_0$ = 1.04. $\langle I^{\mathrm{C}}\rangle$ and $1/L_\text{eff}$ vary with $f_C$, and for these coupler parameters they range from -700 to 700 nA and $1/(1070 ~\mathrm{pH})$ to $1/(-48 ~\mathrm{pH})$, respectively. Note that the slope of $\langle I^{\mathrm{C}}\rangle$ with respect to flux determines the sign of $L_\mathrm{eff}$ and, thus, the sign of $J$. Therefore, Figure \ref{fig:fig2}c can be seen as a map of the regions of anti-ferromagnetic ($J>0$), ferromagnetic ($J<0$), and zero coupling.  

\section{Coupling strength}

\begin{figure}
\begin{center}
\includegraphics[scale=1]{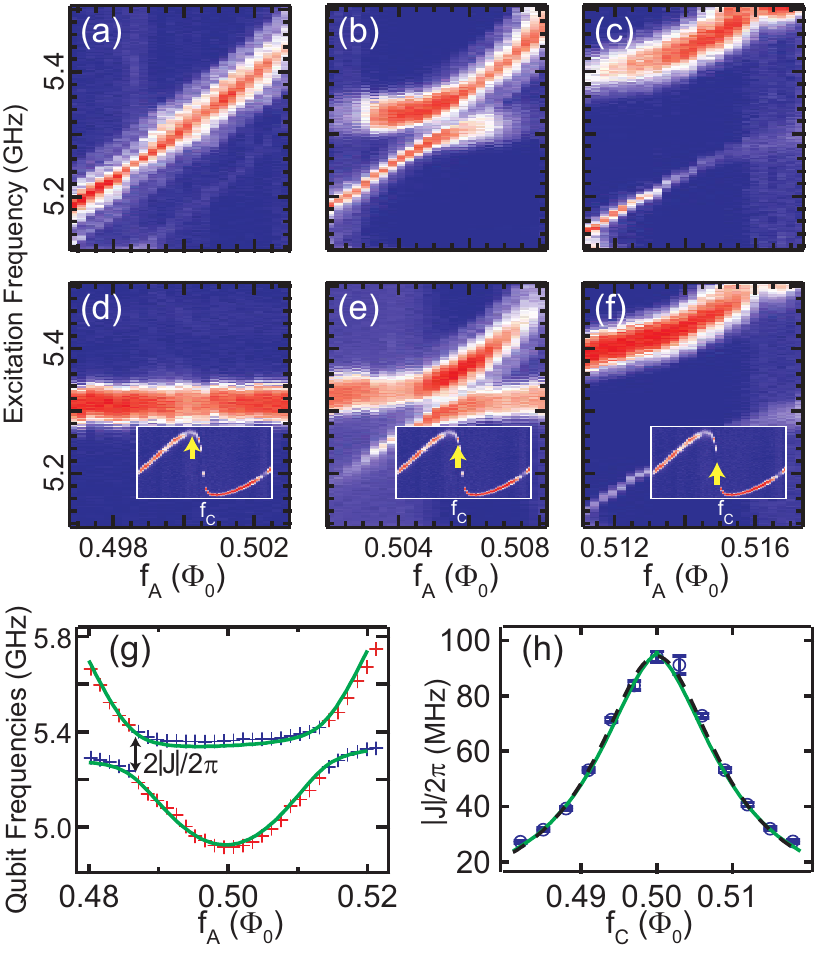}
\caption{\textbf{Qubit-qubit coupling.} Dashed black traces: semi-classical model. Solid green traces: simulations of the full circuit Hamiltonian. (a-f) Spectroscopy of qubit level crossings for different coupling strengths. Panels (a-c) and (d-f) show measurements using Resonator A and Resonator B, respectively. In each panel we scan $f_A$ while holding $f_B$ at a fixed bias point $\sim10~\mathrm{m}\Phi_0$ away from degeneracy.  The left, middle, and right panels correspond to zero ($f_\mathrm{C}=0.402$), intermediate ($f_\mathrm{C}=0.48$), and maximum ($f_\mathrm{C}=0.5$) coupling, as indicated by the insets. (g) Avoided level crossings as Qubit A (red) is tuned across Qubit B (blue) with $f_\mathrm{C} = 0.5$. (h) J vs coupler bias. Error bars are derived from the error of fitting the qubit spectroscopy to a double Gaussian function. }
\label{fig:fig3}
\end{center}
\end{figure}

Two-qubit coupling is shown in Figure \ref{fig:fig3}, focusing primarily on the ferromagnetic coupling regime. Panels (a-f) show spectroscopy of both qubits as the transition frequency of Qubit A is tuned through resonance with that of Qubit B, which is held at a fixed bias.  When the qubits are resonant their levels hybridize and split in frequency by $2|J|$, shown here for three coupler biases corresponding to different values of coupling strength $J$.  Panel (g) shows the qubit frequencies for maximal coupling, as $f_\mathrm{A}$ is tuned over a much larger range.  At this coupler bias, we measure a maximal coupling strength of $|J|/2\pi = 94$ MHz.  From this measurement and our experimental bound on the minimum coupling (see Appendix E), we place a lower bound of $425$ on the coupler on/off ratio.  Finally, panel (h) shows the dependence of $|J|$ on the coupler bias, which agrees well with simulations of the full circuit Hamiltonian, as well as a semi-classical model (see Appendix B).

\section{Qubit coherence}

\begin{figure}
\begin{center}
\includegraphics[scale=1]{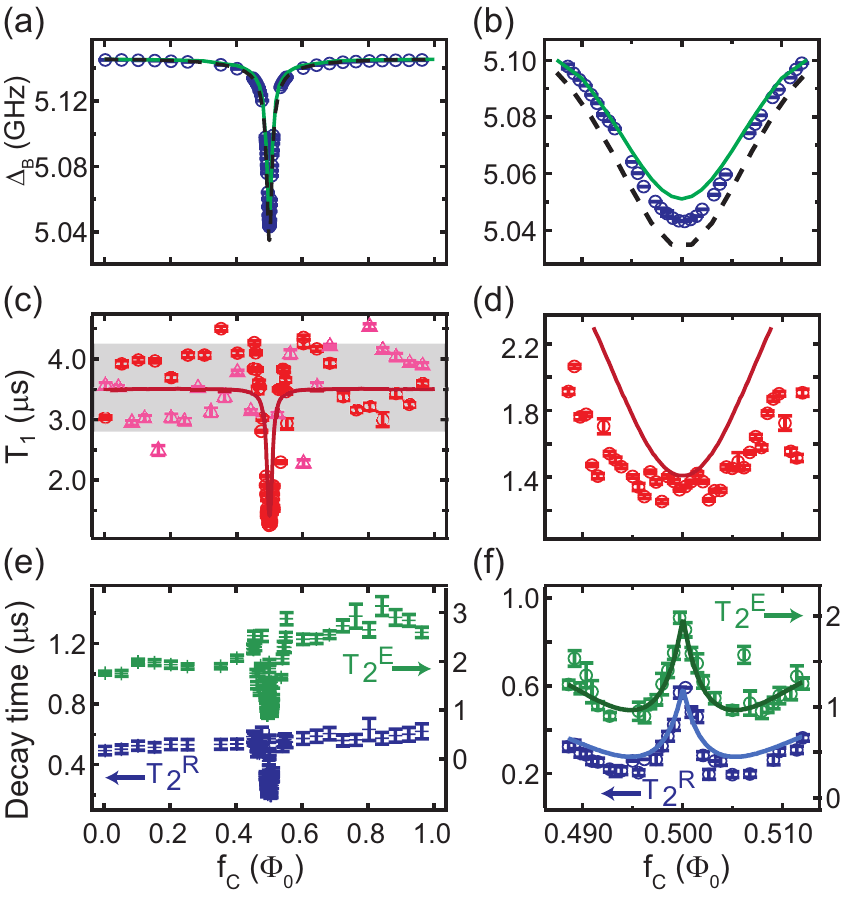}
\caption{\textbf{Qubit properties vs coupler bias.} Results for Qubit B, with $f_\mathrm{A}=0$. Left column: full range of coupler biases. Right column: zoom in near coupler degeneracy. (a,b) $\Delta_\mathrm{B}$ vs coupler bias. Dashed black traces: semi-classical model. Solid green traces: simulations of the full circuit Hamiltonian. (c,d) Qubit energy relaxation time vs coupler bias. The red circles and the magenta triangles correspond to measurements taken at different times. The grey band indicates the typical range of $T_1$ variations when the coupler is biased away from degeneracy. The solid line represents an upper bound on qubit $T_1$ due to flux noise in the coupler loop with an exponent $\alpha=0.91$ and an amplitude of $15~\mu\Phi_0/\sqrt{\mathrm{Hz}}$, combined in parallel with a coupler-independent relaxation time of $3.5~\mu$s. (e,f) Ramsey (left axis) and echo (right axis) $1/e$ decay times vs coupler bias. Solid lines show the expected dependence due to $1/f^\alpha$ flux noise in the coupler loop with the same amplitude and exponent as above.}
\label{fig:fig4}
\end{center}
\end{figure}

In Figure \ref{fig:fig4}, we show how the properties of an individual qubit depend on the coupler bias.  Here, we present data for Qubit B, with $f_\mathrm{A}$ set to zero. Panels (a,b) display $\Delta_\mathrm{B}$ versus $f_\mathrm{C}$.  For each value of $f_\mathrm{C}$, we sweep $f_\mathrm{B}$ and perform qubit spectroscopy to find the minimum qubit frequency, $\omega_\mathrm{B}^\mathrm{min}(f_\mathrm{C})\equiv\Delta_\mathrm{B}(f_\mathrm{C})$. The dependence of $\Delta_\mathrm{B}$ on $f_\mathrm{C}$ can be understood semi-classically as loading of the qubit inductance by the effective inductance of the coupler, 

\begin{equation}
L_\mathrm{q}^{\mathrm{loaded}} = L_\mathrm{q} -  \frac{M^2}{L_\mathrm{eff}},
\end{equation}

\noindent as illustrated by the dashed lines in Figures \ref{fig:fig4}a,b.  

In Figures \ref{fig:fig4}c,d we show how the qubit energy relaxation time $T_1$ depends on $f_\mathrm{C}$.  For each coupler bias point, Qubit B is biased on degeneracy (at the point of minimum qubit frequency). Error bars correspond to the standard error for fitting the decay curve at each coupler bias point to an exponential function. In addition to any dependence on the coupler bias, $T_1$ also fluctuates on slow timescales~\cite{yan16,gust16}, and the grey band indicates the typical range of $T_1$ fluctuations when the coupler is biased away from degeneracy (see Appendix D). When the coupler is biased near degeneracy, we observe a reduction in $T_1$ substantially below the range of temporal fluctuations.

Finally, panels (e,f) show the dependence of the qubit dephasing times on $f_\mathrm{C}$, for the same bias conditions as above. Here, we report the $1/e$ decay times $T_2^{\mathrm{Ramsey}}$ and $T_2^{\mathrm{Echo}}$ for Ramsey interferometry and spin echo experiments, respectively.  When the coupler is biased away from degeneracy, $T_2^{\mathrm{Ramsey}}$ is essentially constant with respect to $f_\mathrm{C}$.  There is some variation in $T_2^{\mathrm{Echo}}$, which is roughly consistent with the range of  values expected from the observed fluctuations in $T_1$. 

Interestingly, we observe a sharp reduction in the coherence times as the coupler bias approaches degeneracy, and a full recovery when the coupler is biased exactly on degeneracy.  This effect can be understood as the result of the first-order sensitivity of $\Delta_\mathrm{B}$ to the coupler bias, which is given by $\partial\Delta_\mathrm{B}/\partial\Phi_\mathrm{C}$, the slope of the data in panels (a,b).  By fitting the measured dependence of $\Delta_\mathrm{B}$ on $f_\mathrm{C}$ and assuming a $1/f^{\alpha}$ spectral density of fluctuations with $\alpha =  0.91$~\footnote{In our previous work~\cite{yan16}, we determined that $\alpha = 0.9$ for capacitively shunted flux qubits with $10\times10$ $\mu\text{m}^2$ loops produced using our fabrication process.  Here, we have chosen to use $\alpha = 0.91$ in order achieve a good fit to our coherence data, as discussed in Appendix D.}, we see excellent agreement between our model and the coherence measurements for a flux noise amplitude of $15$ $\mu\Phi_0/\sqrt{\text{Hz}}$, as indicated by the curves in Figure 4f. Using the same amplitude and exponent, we calculate an upper limit on qubit $T_1$ due to flux noise in the coupler loop, as shown in Figure 4c,d.  In Appendix D, we speculate on why the estimated flux noise amplitude is larger than previously reported values for flux qubits made with the same fabrication process~\cite{yan16} and the potential implications for future quantum annealing architectures designed to optimize for both coherence and coupling.

This work represents an important step toward designing quantum annealers with improved coherence. We have demonstrated tunable coupling between flux qubits with substantially lower persistent currents than existing commercial devices, thereby reducing the qubit sensitivity to flux noise in their respective loops.  This approach requires an increased coupler susceptibility, which increases the qubits' sensitivity to flux noise in the coupler loop.  We have examined this effect by measuring qubit coherence across the full range of coupler biases, using standard measurement techniques borrowed from the gate-based quantum computing community, which have yet to be applied to commercial quantum annealers.  Looking forward, our approach can be extended to achieve larger coupling strength, symmetric bipolar coupling, and $\hat{\sigma}_\mathrm{x}\hat{\sigma}_\mathrm{x}$ interactions~\cite{kerm08}, while maintaining low persistent currents.  Our results provide new insights into the available design space and suggest the type of systems-level analysis that will be necessary when designing quantum annealers with improved coherence.  

\begin{table*}
\centering
\begin{tabular}{ |c|l|c|c| }
\hline
& ~~Parameter & ~Semi-classical model~ & ~Full galvanic circuit model~ \\ \hhline{|=|=|=|=|}
\multirow{2}{*}{Common junction params.~} & ~$J_\mathrm{c}$ ($\mu$A)& 2.78 & 2.78 \\
 & ~$S_\mathrm{c}$ (fF/$\mu \mathrm{m}^2$)~ & 50 & 50 \\ \hline
\multirow{4}{*}{Qubit A} & ~$I_\mathrm{0}^{\mathrm{A,sm}} (\mathrm{nA})$ & 78 & 78 \\
& ~$I_\mathrm{0}^{\mathrm{A,lg}} (\mathrm{nA})$ & 206 & 206 \\
& ~$C_\mathrm{sh}^{\mathrm{A}}$ (fF) & 53 & 53 \\
& ~$L_\mathrm{q}^\mathrm{A}$ (pH) & 115 & 115 \\ \hline
\multirow{4}{*}{Qubit B} & ~$I_\mathrm{0}^{\mathrm{B,sm}} (\mathrm{nA})$ & 78 & 78 \\
& ~$I_\mathrm{0}^{\mathrm{B,lg}} (\mathrm{nA})$ & 209 & 209 \\
& ~$C_\mathrm{sh}^{\mathrm{B}}$ (fF) & 53 & 53  \\
& ~$L_\mathrm{q}^\mathrm{B}$ (pH) & 115 & 115 \\ \hline
\multirow{3}{*}{Coupling} & ~$M$ (pH) & 39 & 43 \\
&~$I_\mathrm{0}^{\mathrm{C}} (\mathrm{nA})$ & 727 & 736 \\
&~$L_\mathrm{C}$ (pH) & 467 & 542 \\  \hline
\end{tabular}
\caption{\label{tab1} Table of sample parameters.}
\end{table*}

\begin{acknowledgments}
We gratefully acknowledge Wayne Woods for useful discussions and P. Baldo, G. Fitch, X. Miloshi, P. Murphy, B. Osadchy, K. Parrillo, A. Sevi, R. Slattery at MIT Lincoln Laboratory for technical assistance. This research was funded by the Office of the Director of National Intelligence (ODNI), Intelligence Advanced Research Projects Activity (IARPA) and by the Assistant Secretary of Defense for Research \& Engineering under Air Force Contract No. FA8721-05-C-0002. The views and conclusions contained herein are those of the authors and should not be interpreted as necessarily representing the official policies or endorsements, either expressed or implied, of ODNI, IARPA, or the US Government.
\end{acknowledgments}

\appendix

\section{Table of sample parameters}

\label{params}

Table 1 shows a list of sample parameters extracted from two different models of the coupled qubit system\textemdash  a semi-classical model, where the individual qubits and coupler are treated quantum mechanically but their interactions are treated as a classical mutual inductance, and a quantum model of the full galvanically-coupled circuit.  Using the semi-classical model, we performed an optimization routine to determine the set of parameters which best fit the results in Figures 2-4.  The quantum model includes some effects, such as cross-capacitance between the qubits and coupler, which are not included in the semi-classical model.  Therefore, it was necessary to make small adjustments to the parameters extracted from the semi-classical model in order to achieve good agreement between the quantum model and the measured results, as indicated in Table 1.

\section{Semi-classical model}

\begin{figure*}
\begin{center}
\includegraphics[scale=1]{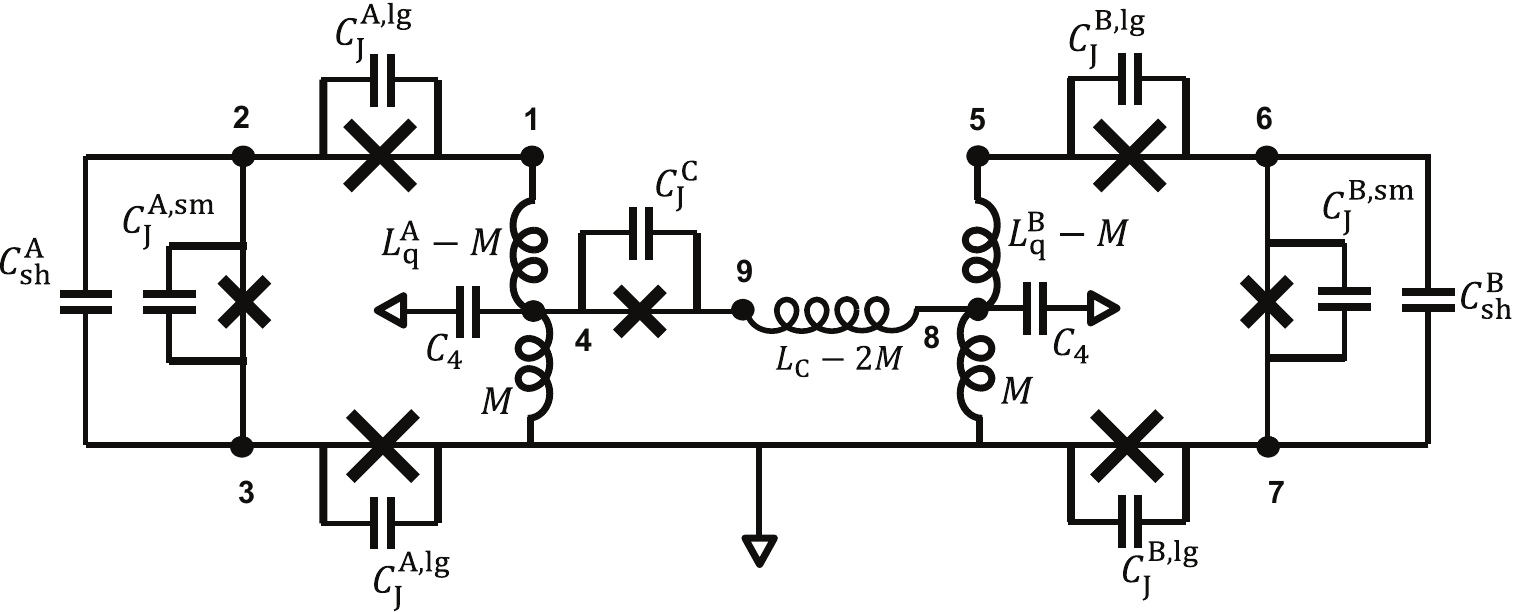}
\caption{\textbf{Schematic diagram of the full galvanic circuit.} The nodes of the circuit, labeled $1-9$, are used to define its canonical flux and charge variables~\cite{devo03}.}
\label{fig:FullCir}
\end{center}
\end{figure*}

Figure \ref{fig:FullCir} shows a circuit diagram for the full galvanically-coupled circuit. To simulate the energy levels of the full system, we diagonalize the circuit Hamiltonian, using similar techniques to our previous work~\cite{yan16}.  These techniques are described in detail in a separate forthcoming publication~\footnote{A. J. Kerman, in preparation.}.

In this section, we explain how to map the full circuit onto a simpler and more computationally convenient semi-classical model.  In this model, the individual qubits and coupler are treated quantum mechanically, but their interactions are treated as a classical mutual inductance.  Using this simplified model, we derive expressions for the coupling strength $J$, as well as the shifts in the qubit parameters $\Delta$ and $\epsilon$ due to interaction with the coupler.  

\subsection{Comparing mutually-inductive coupling to galvanic coupling}

\begin{figure*}
\begin{center}
\includegraphics[scale=1]{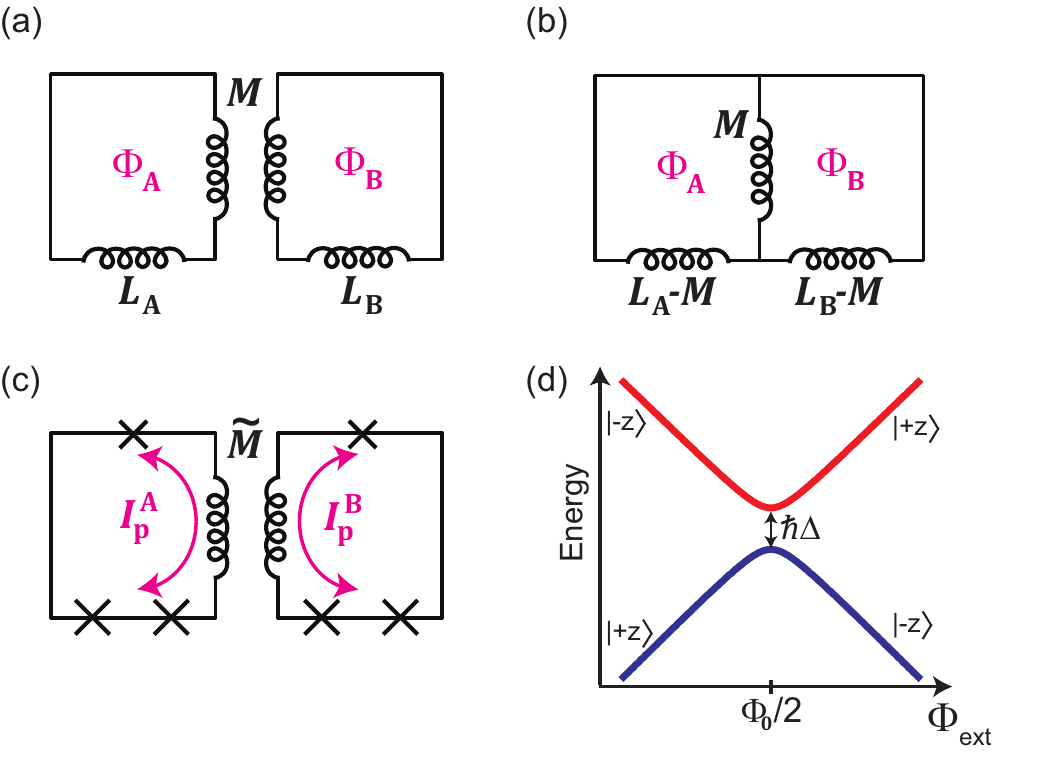}
\caption{\textbf{Direct coupling.} (a) Circuit schematic for two loops of inductance $L_\mathrm{A,B}$ coupled through a mutual inductance $M$. (b) Circuit schematic for two loops which are galvanically coupled through a shared inductance $M$.  (c) Circuit schematic for two flux qubits with persistent currents $I_\mathrm{p}^\mathrm{A,B}$ coupled through a mutual inductance $\tilde{M}$. (d) Illustration of the energies of the ground (blue) and first-excited (red) states of a flux qubit as a function of the external flux $\Phi_\mathrm{ext}$ through the qubit loop.  At the degeneracy point ($\Phi_\mathrm{ext} = \Phi_0/2$), the ground and excited states are separated in energy by $\hbar \Delta$.  When biased away from degeneracy, the qubit states are approximately persistent current states $\ket{\pm z}$.}
\label{fig:figS1}
\end{center}
\end{figure*}

To build up the model of the coupled qubit system, we first consider a simpler system depicted in Figure~\ref{fig:figS1}a.  Here, two loops of inductance $L_\mathrm{A,B}$ threaded by magnetic flux $\Phi_\mathrm{A,B}$ are coupled through a mutual inductance $M$.  

Defining the flux vector $\bm{\Phi}$, the mutual inductance matrix $\bm{M}$, and the self-inductance matrix $\bm{L}$ as

\begin{equation}
\bm{\Phi} \equiv \begin{pmatrix} \Phi_\mathrm{A} \\ \Phi_\mathrm{B} \end{pmatrix};~\bm{M} \equiv \begin{pmatrix} 0&-M \\ -M&0 \end{pmatrix};~\bm{L} \equiv \begin{pmatrix} L_\mathrm{A}&0 \\ 0&L_\mathrm{B} \end{pmatrix},
\end{equation}

\noindent the classical potential energy of the system is given by

\begin{align}
U &= \frac{1}{2} \bm{\Phi}\left(\bm{L}^{-1}+\bm{L}^{-1}\bm{M}\bm{L}^{-1}\right)\bm{\Phi} \\ &=  \frac{1}{2} \frac{\Phi_\mathrm{A}^2}{L_\mathrm{A}} +\frac{1}{2} \frac{\Phi_\mathrm{B}^2}{L_\mathrm{B}} + M \frac{\Phi_\mathrm{A}}{L_\mathrm{A}}\frac{\Phi_\mathrm{B}}{L_\mathrm{B}}, \label{eqU}
\end{align}

\noindent where the first two terms correspond to the energies of the individual loops, and the third term represents their interaction energy.  The system can be re-expressed in terms of the classical circulating currents $I_\mathrm{A,B} = \Phi_\mathrm{A,B}/L_\mathrm{A,B}$, which yields

\begin{equation}
\label{eqS1}
U =\frac{1}{2} L_\mathrm{A}I_\mathrm{A}^2 + \frac{1}{2} L_\mathrm{B}I_\mathrm{B}^2 + MI_\mathrm{A}I_\mathrm{B}.
\end{equation}

Next, we will compare this result for two mutually coupled loops to the case of two galvanically coupled loops, as depicted in Figure \ref{fig:figS1}b.  Here, the inductance matrix can be approximately defined as~\footnote{This 2x2 inductance matrix is appropriate in the limit of small island capacitance of the node connecting the three inductors; In this limit, independent phase fluctuations of this node can be neglected, and the circuit can be described by only two canonical phase variables with a 2x2 inductance matrix.}

\begin{equation}
\bm{L} \equiv \begin{pmatrix} L_\mathrm{A}&-M \\ -M&L_\mathrm{B} \end{pmatrix}
\end{equation}

\noindent and the potential energy is given by

\begin{align}
U &= \frac{1}{2} \bm{\Phi}\bm{L}^{-1}\bm{\Phi}\\
 &= \frac{1}{2} \frac{\Phi_\mathrm{A}^2}{L_\mathrm{A}-M^2/L_\mathrm{B}} + \frac{\Phi_\mathrm{B}^2}{L_\mathrm{B}-M^2/L_\mathrm{B}} + M\left(1-\frac{M^2}{L_\mathrm{A}L_\mathrm{B}}\right)^{-1} \frac{\Phi_\mathrm{A}}{L_\mathrm{A}}\frac{\Phi_\mathrm{B}}{L_\mathrm{B}}.
\end{align}

\noindent Note that this is equivalent to equation~\eqref{eqU} after the following substitutions:

\begin{equation}
\label{galvanicSub}
\tilde{L}_\mathrm{A,B} \equiv L_\mathrm{A,B} - \frac{M^2}{L_\mathrm{B,A}};~\tilde{M}\equiv M\left(1-\frac{M^2}{L_\mathrm{A}L_\mathrm{B}}\right).
\end{equation}

\noindent Thus, the galvanically-coupled circuits employed in this work  can be approximately mapped onto simpler mutually-coupled circuits using the renormalized inductances  $\tilde{L}$ and $\tilde{M}$.

\subsection{Directly-coupled qubits}

Now, suppose that each loop in the circuits discussed above is replaced with a flux qubit (FIG.~\ref{fig:figS1}c,d) described by the Hamiltonian $H_\mathrm{q}/ \hbar \approx (\epsilon\hat{\sigma}_{\mathrm{z}}+\Delta\hat{\sigma}_{\mathrm{x}})/2$.  In the persistent current basis, the eigenstates of the Pauli operator $\hat{\sigma}_\mathrm{z}$, denoted $\ket{\pm z}$, correspond to  clockwise and counter-clockwise circulating currents 

\begin{equation}
\label{eqS2}
I \equiv \bra{\pm z}\hat{I}\ket{\pm z} = \bra{\pm z}I_\mathrm{p}\hat{\sigma}_{\mathrm{z}}\ket{\pm z}= \pm I_\mathrm{p},
\end{equation}

\noindent where $\hat{I}$ is the current operator and $I_\mathrm{p}$ is magnitude of the qubit persistent current.  The interaction term from equation~\eqref{eqS1} can be expressed as

\begin{equation}
\label{eqS3}
H_\mathrm{int} = \tilde{M}I_\mathrm{p}^\mathrm{A}I_\mathrm{p}^\mathrm{B}\hat{\sigma}_{\mathrm{z}}^{(\mathrm{A})}\hat{\sigma}_{\mathrm{z}}^{(\mathrm{B})},
\end{equation}

\noindent which takes the form $H_{\mathrm{int}}=J\hat{\sigma}_{\mathrm{z}}^{(\mathrm{A})}\hat{\sigma}_{\mathrm{z}}^{(\mathrm{B})}$, where the coupling strength $J$ is given by

\begin{equation}
\label{eqS4}
\hbar J = \tilde{M}I_\mathrm{p}^\mathrm{A}I_\mathrm{p}^\mathrm{B}.
\end{equation}

A simple intuitive picture for this expression emerges when the qubits are biased such that $\epsilon\gg\Delta$. In this regime, qubit energy eigenstates are approximately equal to the persistent current states $\ket{\pm z}$ with energy eigenvalues  $\pm \hbar \epsilon/2 = I_\mathrm{p}(\Phi_\mathrm{ext}-\Phi_0/2)$, where $\Phi_\mathrm{ext}$ is the external magnetic flux through the qubit loop and $\Phi_0$ is the magnetic flux quantum. Here, the $\hat{\sigma}_{\mathrm{z}}\hat{\sigma}_{\mathrm{z}}$ interaction is longitudinal with respect to the energy eigenbasis. The coupling can be understood by considering the effect of the persistent current in one qubit loop on the flux through the other qubit loop.  For example, Qubit A induces a state-dependent offset $\delta \Phi_\mathrm{B} = \pm \tilde{M} I_\mathrm{p}^\mathrm{A}$ in the flux through Qubit B and thus a state-dependent frequency shift of 

\begin{equation}
\label{eqS5}
\delta \omega_\mathrm{01}^{(B)} \approx \pm \epsilon \delta \Phi_\mathrm{B} = \pm 2J.
\end{equation}

Note that the coupling measurements reported in the main text were performed in the $\Delta\gg\epsilon$ regime, where the $\hat{\sigma}_{\mathrm{z}}\hat{\sigma}_{\mathrm{z}}$ interaction is transverse with respect to the energy eigenbasis.  In this case, the coupling manifests as an avoided crossing between the $\ket{01}$ and $\ket{10}$ states, which are shifted from their bare energies by $\pm \hbar J$.  

\subsection{Mediated coupling}

\begin{figure*}
\begin{center}
\includegraphics[scale=1]{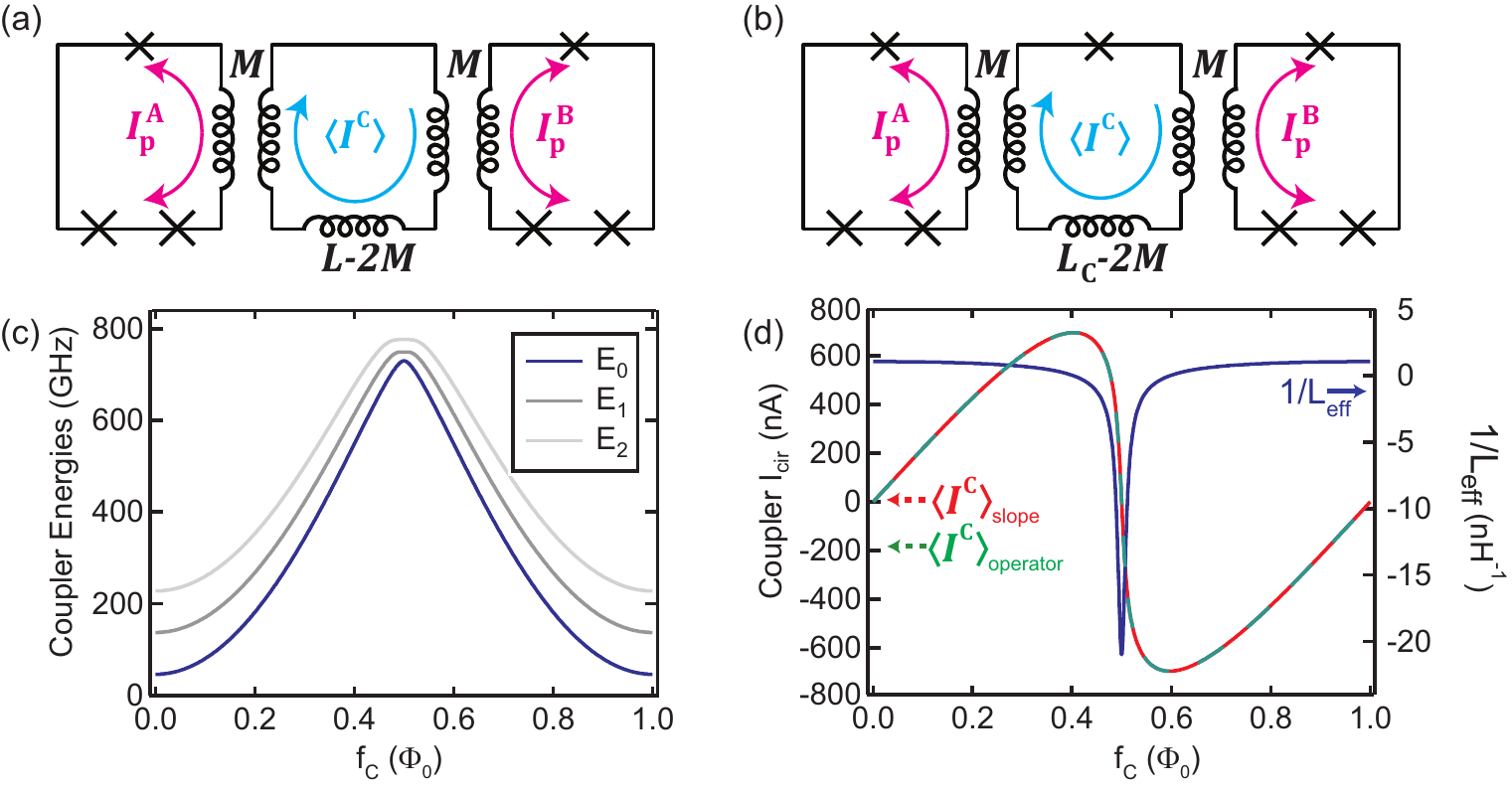}
\caption{\textbf{Mediated coupling.} (a) Circuit schematic for two flux qubits with persistent currents $I_\mathrm{p}^\mathrm{A,B}$ which each couple through a mutual inductance $\tilde{M}$ to an intermediate loop of inductance $L$. (b) Circuit schematic for a similar configuration, but with the intermediate loop replaced with an RF-SQUID coupler. (c) Illustration of the energies of the ground (blue) and first-excited (grey) states of a an RF-SQUID coupler, as a function of the external flux $\Phi_\mathrm{c}$ through the coupler loop.  (d) Left axis: comparison of the coupler circulating current, calculated using the slope of the ground state energy (red) and using  and using the current operator (green). Right axis: effective inductance of the coupler vs. $\Phi_\mathrm{c}$.}
\label{fig:figS2}
\end{center}
\end{figure*}

As a next step in building up the semi-classical coupling model, we consider the case of two qubits coupled through a mutual inductance $\tilde{M}$ to an intermediate loop of  inductance $L$ (FIG.~\ref{fig:figS2}a).  Returning to the longitudinal coupling picture ($\epsilon \gg \Delta$), the persistent current in Qubit A will induce a state-dependent offset flux $\delta \Phi_\mathrm{C}=\pm \tilde{M} I_\mathrm{p}^\mathrm{A}$ in the coupler loop, which changes the current circulating in the loop by $\delta \langle I^{\mathrm{C}}\rangle = \delta \Phi_\mathrm{C}/L$ and thus induces an offset of 

\begin{equation}
\label{eqS6}
\delta \Phi_\mathrm{B} = \delta \Phi_\mathrm{C} \frac{\tilde{M}}{L} = \frac{\tilde{M}^2}{L} I_\mathrm{p}^\mathrm{A}
\end{equation}

\noindent in the flux through Qubit B.  Note that this expression takes the same form as for the directly-coupled qubits, but with the substitution $\tilde{M} \rightarrow \tilde{M}^2/L\equiv M_\mathrm{eff}$.  Then, in analogy to equation~\eqref{eqS4}, the coupling strength is given by 

\begin{equation}
\label{eqS7}
J = \frac{\tilde{M}^2}{L}I_\mathrm{p}^\mathrm{A}I_\mathrm{p}^\mathrm{B} = M_\mathrm{eff}I_\mathrm{p}^\mathrm{A}I_\mathrm{p}^\mathrm{B}.
\end{equation}

Finally, we consider the case where the intermediate loop is replaced with an RF-SQUID coupler (FIG.~\ref{fig:figS2}b).  In the following discussion, we make the assumption that transition frequency between the coupler ground- and first-excited state is much larger than the qubit frequencies and that the coupler is always operated in its ground state.  In general, the coupler ground state energy $E_0$ varies with applied flux $f_\mathrm{C}$, as illustrated in Figure~\ref{fig:figS2}c.  For the coupler parameters considered in this work, the circulating current in the coupler loop is approximately equal to the slope of coupler energy with respect to flux, $\langle I^{\mathrm{C}}\rangle\approx \partial E_0^\mathrm{C}/ \partial\Phi_\mathrm{C}$, as illustrated in Figure ~\ref{fig:figS2}d, where we compare this quantity with the expectation value of the current operator $\bra{g}\hat{I}\ket{g}$ for the coupler ground sate $\ket{g}$.

We then define the ``quantum inductance" for the coupler (as in References~\cite{joha06,torn07,harr09} and in analogy to the ``quantum capacitance" described in the charge qubit~\cite{sill05,duty05,joha06} and semi-conducting qubit~\cite{coll13, pete10, ilan06, glaz92,fogl05, late05} literature) as

\begin{equation}
\frac{1}{L_\mathrm{eff}} \equiv \frac{\partial \langle I^{\mathrm{C}}\rangle}{\partial\Phi_\mathrm{C}} \approx \frac{\partial^2 E_0^{(\mathrm{C})}}{\partial\Phi_\mathrm{C}^2}.
\end{equation}

\noindent Note that unlike a physical inductance, this quantum inductance can take both positive and negative values.  Following the same logic as above, we can now express the coupling strength as

\begin{equation}
\label{J}
J = \frac{\tilde{M}^2}{L_\mathrm{eff}}I_\mathrm{p}^\mathrm{A}I_\mathrm{p}^\mathrm{B}.
\end{equation}

Given a set of qubit and coupler parameters, it is straightforward to calculate $J$ using equation~\eqref{J}.  We determine $L_\mathrm{eff}$ by numerically diagonalizing the coupler Hamiltonian to solve for its ground state energy $E_0$ as a function of $\Phi_\mathrm{C}$.  We separately determine $I_\mathrm{p}$ by numerically solving for the energy eigenstates $\ket{\psi_j}$ of the qubit Hamiltonian, from which we calculate the matrix elements of the current operator, $\bra{\psi_j}\hat{I}\ket{\psi_k}$, expressed in the energy eigenbasis.  When the qubit is biased on degeneracy ($\epsilon = 0$), the $I_\mathrm{p}$ is given by the off-diagonal matrix elements.  

Note that  equation~\eqref{J} is the same expression for coupling strength used by D-Wave in references~\cite{harr07,harr09}, with the coupler susceptibility $\chi$ defined as the inverse of the effective inductance. However, their approach differs in that instead of diagnolizing the coupler Hamiltonian to solve for $\chi$, D-Wave chooses to approximate $\chi$ as the first-order (linear) susceptibility, which can be expressed using a simple analytic formula.  This approach works sufficiently well for the coupler parameters of existing D-Wave devices, but the linear approximation breaks down for larger coupler susceptibilities and coupling strengths, as discussed in reference~\cite{kafr16}.  

\subsection{Qubit flux offset due to coupler}

The semi-classical model can also explain the shifts in qubit parameters due to their interaction with the coupler.  For concreteness and to follow the presentation of the main text, we will focus on Qubit B.  First, we consider the effect of the coupler on the qubit flux bias.  This effect explains the dependence of the qubit frequency on the coupler bias shown in Figure 2c.  

As shown in Figure ~\ref{fig:figS2}d, the circulating current in the coupler loop $\langle I^{\mathrm{C}}\rangle$ varies with the coupler bias $\Phi_\mathrm{C}$.  This circulating current couples into the qubit loop through the mutual inductance $M$, and therefore threads a flux

\begin{equation}
\delta \Phi_\mathrm{B} = \tilde{M} \langle I^{\mathrm{C}}\rangle
\end{equation}

\noindent through the qubit loop.  For a flux qubit described by the Hamiltonian $H_\mathrm{q}/ \hbar \approx (\epsilon\hat{\sigma}_{\mathrm{z}}+\Delta\hat{\sigma}_{\mathrm{x}})/2$, this flux offset corresponds to a shift in $\epsilon$ of

\begin{equation}
\delta \epsilon = \frac{2}{\hbar}I_\mathrm{p}\delta\Phi_\mathrm{B} = \frac{2}{\hbar}\tilde{M}I_\mathrm{p}\langle I^{\mathrm{C}}\rangle.
\end{equation}

\subsection{Inductive loading model}

The coupler also affects the value of $\Delta$, the qubit frequency when biased at its degeneracy point, as shown for Qubit B in Figure 4a,b.  This effect can be modeled semi-classically as inductive loading of the qubit inductance by the effective inductance of the coupler.  

\begin{figure}
\begin{center}
\includegraphics[scale=1]{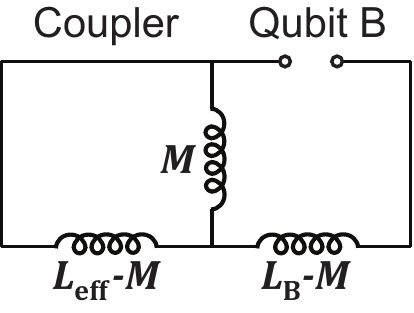}
\caption{\textbf{Inductive loading model.} Circuit schematic used to model the loading of the qubit inductance due to the coupler $L_\mathrm{eff}$.}
\label{fig:figS3}
\end{center}
\end{figure}

A circuit schematic for the inductive loading model is shown in Figure~\ref{fig:figS3}~\footnote{Note that a general treatment would also include the effective inductance of Qubit A, $L_\mathrm{eff}^\mathrm{(A)}$, but for the device parameters presented here, $L_\mathrm{eff}^\mathrm{(A)}\gg L_\mathrm{eff}^\mathrm{(C)}$ and therefore has a negligible effect on Qubit B.}.  Here, we consider the impedance looking out from the Josephson junction, to calculate loaded qubit inductance

\begin{equation}
\label{LoadedL}
L_\mathrm{B}^\mathrm{loaded} = L_\mathrm{B}-M+\left(\frac{1}{L_\mathrm{eff}-M} + \frac{1}{M} \right)^{-1} = L_\mathrm{B}-\frac{M^2}{L_\mathrm{eff}}.
\end{equation}

\noindent Note that this expression for the loaded inductance is the same as the renormalized inductance derived in equation~\eqref{galvanicSub}.  To calculate the semi-classical theory curves for $\Delta_\mathrm{B}$ versus $f_\mathrm{C}$ (FIG. 4a,b), we first simulate the coupler to determine $L_\mathrm{eff}(f_\mathrm{C})$ (FIG. ~\ref{fig:figS2}d). Then, for each value of $f_\mathrm{C}$, we determine $\Delta_\mathrm{B}$ by simulating the qubit energy levels using $L_\mathrm{B}^\mathrm{loaded}$ for the qubit loop inductance. 

\section{Variations of $T_1$ in time}

In addition to any systematic dependence of $T_1$ on the qubit and coupler biases, we also observe $T_1$ variations in time.  These variations are possibly related to quasiparticle fluctuations, as described in reference~\cite{gust16}.  Although $T_1$ fluctuations are not a primary focus of this work, they affect the interpretation of the data shown in Figure 4c,d.  In Figure 9 we show the results of repeated $T_1$ measurements for Qubit B, with the coupler biased away from degeneracy.  The range of observed $T_1$ values over the $14$ hour measurement time is represented as a grey band in Figure 4c.  

\begin{figure*}
\begin{center}
\includegraphics[scale=1]{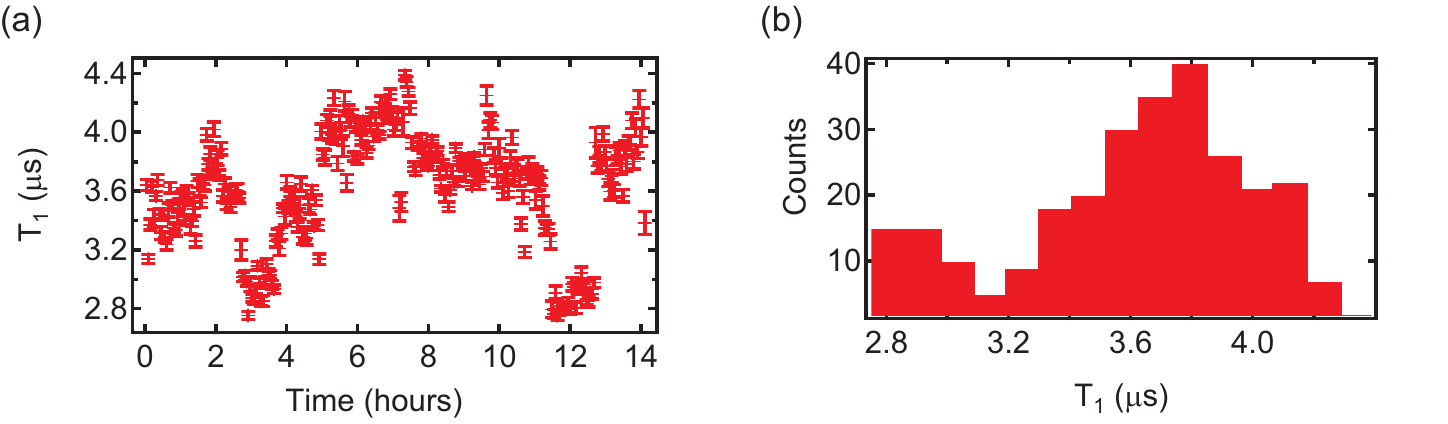}
\caption{\textbf{Repeated $T_1$ measurements.} (a) Repeated measurements of $T_1$ of Qubit B as a function of time. Here, $f_\mathrm{A} = 0$, $f_\mathrm{B} = 0.5$, and $f_\mathrm{C} = 0$. (b) Histogram of measured $T_1$ values.  }
\end{center}
\label{fig:T1}
\end{figure*}

\section{Modeling the effect of flux noise on qubit coherence}

\subsection{Definition of noise spectral density}

In this work, as in reference~\cite{yan16}, we choose to characterize noise by the symmetric power spectral density (PSD)

\begin{equation}
S_\lambda(\omega) = \int_{-\infty}^{\infty}\mathrm{d}\tau \mathrm{exp}(-i\omega \tau)\frac{1}{2}\langle \hat{\lambda}(0) \hat{\lambda}(\tau)+ \hat{\lambda}(\tau) \hat{\lambda}(0)\rangle,
\end{equation}

\noindent where $\hat{\lambda}$ is an operator representing a fluctuating parameter $\lambda$.  The two dominant noise mechanisms for the coupled qubit system presented here are flux noise in the qubit loop and the coupler loop, $\lambda_i = \Phi_\mathrm{B},\Phi_\mathrm{C}$. For $1/f$-like noise, the noise amplitude $A_\lambda$ is given by~\footnote{Note that our definition for $S_\lambda(\omega)$ differs by a factor of $2\pi$ from the expression used in reference~\cite{byla11}, but for the case of $\gamma=1$ our definitions of $A_\lambda$ are equivalent. Also note that our definition of $S_\lambda(\omega)$ is double-sided, and thus differs by a factor of $2$ from the single-sided spectral density used in reference~\cite{sank12}.}

\begin{equation}
\label{A}
S_\lambda(\omega) = A_\lambda^2\left(\frac{2\pi \times 1 \mathrm{Hz}}{\omega}\right)^\gamma,
\end{equation}

\noindent where $\gamma\sim 1$.

\subsection{Energy relaxation due to $1/f^\gamma$-flux noise}

We have analyzed the data for $T_1$ of our qubit-coupler system using the Fermi's golden rule model presented in reference~\cite{yan16},

\begin{equation}
\frac{1}{T_1}=\sum_\lambda2\frac{|\bra{e}\hat{D}_\lambda\ket{g}|^2}{\hbar^2}S_\lambda\left(\omega\right),
\end{equation}

\noindent where the sum is taken over decay mechanisms, $S_\lambda\left(\omega\right)$ is the power spectral density of the noise responsible for each decay mechanism, and the operator $\hat{D}_\lambda$ is the transition dipole moment which couples our system to each noise source.

For the coupled system considered here, $T_1$ can be decomposed into contributions from the qubit, $T_1^\mathrm{Q}$, and the coupler, $T_1^\mathrm{C}$, where

\begin{equation}
\frac{1}{T_1}=\frac{1}{T_1^\mathrm{Q}}+\frac{1}{T_1^\mathrm{C}}.
\end{equation}

\noindent The qubit contribution  dominates away from coupler degeneracy, and both processes contribute when the system is biased near coupler degeneracy.

In our analysis, we assume that the coupler is flux noise limited on its degeneracy, and its decay rate is thus given by

\begin{equation}
\label{T1C}
\frac{1}{T_1^\mathrm{C}}=2\frac{|\bra{e}\hat{I}^\mathrm{C}\ket{g}|^2}{\hbar^2}S_{\Phi_\mathrm{C}}\left(\omega\right),
\end{equation}

\noindent where $\ket{g}$ and $\ket{e}$ are the ground and first excited state of the coupled system

\begin{equation}
\hat{H}=\hat{H}^\mathrm{Q}+\hat{H}^\mathrm{C}+M\hat{I}^\mathrm{Q}\hat{I}^\mathrm{C}.
\end{equation}

\noindent The quantum operators $\hat{H}^\mathrm{C}$ and $\hat{I}^\mathrm{C}$ ($\hat{H}^\mathrm{Q}$ and $\hat{I}^\mathrm{Q}$) are the Hamiltonian and loop current operator of the bare coupler (qubit) respectively, and the exact value of the matrix element $\bra{e}\hat{I}^\mathrm{C}\ket{g}$ can thus be calculated from the device parameters listed in Section~\ref{params} and the full quantum model of the bare qubit and coupler. As described in Section E, the amplitude and exponent of the flux noise power spectral density in our coupler loop are then chosen to fit the measured values of $T_1$, $T_2^{\mathrm{Ramsey}}$, and $T_2^\mathrm{Echo}$ on coupler degeneracy.

\subsection{First order sensitivity to flux noise}

The sensitivity $\kappa_\lambda$ of the qubit freqency to a parameter $\lambda$ determines the effect of fluctuations in $\lambda$ on qubit dephasing.  In the two-level approximation of the flux qubit, the qubit transition frequency is given by $\omega_\mathrm{01} \approx \sqrt{\epsilon^2+\Delta^2}$, and, to first order,

\begin{equation}
\kappa_\lambda \equiv\frac{\partial \omega_\mathrm{01}}{\partial \lambda} \approx \frac{\partial\omega_\mathrm{01}}{\partial\epsilon}\frac{\partial \epsilon}{\partial \lambda} + \frac{\partial\omega_\mathrm{01}}{\partial\Delta}\frac{\partial \Delta}{\partial \lambda} = \frac{\epsilon}{\omega_{\mathrm{01}}} \kappa_{\epsilon, \lambda} + \frac{\Delta}{\omega_{\mathrm{01}}} \kappa_{\Delta, \lambda},
\end{equation}

\noindent where $\kappa_{\epsilon, \lambda} \equiv \partial \epsilon/\partial \lambda$ and $\kappa_{\Delta, \lambda} \equiv \partial \Delta/\partial \lambda$.

In the measurements presented in Figure 4, we characterized the coherence of Qubit B when biased near its degeneracy point ($\epsilon_\mathrm{B} = 0$).  At this bias point, $\kappa_{\epsilon_\mathrm{B}, \Phi_\mathrm{B}}$ and $\kappa_{\epsilon_\mathrm{B}, \Phi_\mathrm{C}}$ are zero.  Since $\Delta_\mathrm{B}$ depends only weakly on $\Phi_\mathrm{B}$, the dominant first-order noise mechanism is $\kappa_{\Delta_\mathrm{B},\Phi_\mathrm{C}}$, the sensitivity of $\Delta_\mathrm{B}$ to the coupler flux.

\subsection{Decoherence due to $1/f^\gamma$-flux noise}

Here, we consider the effect of $1/f$-like flux noise, as defined in equation~\eqref{A}, on qubit coherence.  In general, this type of noise causes phase decay of the form  $\mathrm{exp}[-\chi_N(t)]$, where~\cite{byla11}

\begin{equation}
\label{chi}
\chi_N(\tau) = \frac{1}{2\pi}\tau^2 \sum_{\lambda}\kappa_\lambda^2\int_0^\infty \mathrm{d}\omega S_\lambda(\omega)g_N(\omega,\tau),
\end{equation}

\noindent where $\tau$ is the free evolution time and $g_N$ is a filter function which depends on the qubit pulse sequence.  For the Ramsey ($N=0$) and a Hanh echo sequences ($N=1$) considered in this work,

\begin{align}
g_0(\omega,\tau) &\equiv g_0(\omega\tau)= \left(\frac{\mathrm{sin}(\omega \tau/2)}{(\omega \tau/2)}\right)^2,\\
g_1(\omega,\tau) &\equiv g_1(\omega\tau)= \left(\frac{\mathrm{sin}(\omega \tau/4)}{(\omega \tau/4)}\right)^2\mathrm{sin}^2(\omega\tau/4).
\end{align}

Substituting equation~\eqref{A} into equation~\eqref{chi} and making the additional substitution $\omega \tau \rightarrow z$ gives

\begin{equation}
\chi_N(\tau) = \frac{(2\pi\times 1~\mathrm{Hz})^\gamma}{2\pi}\tau^{1+\gamma}\sum_{\lambda}\kappa_\lambda^2 A_\lambda^2 \int_0^\infty \frac{\mathrm{d}z}{z^\gamma}g_N(z),
\end{equation}

\noindent where we have assumed that the fluctuations in each parameter $\lambda$ share a common noise exponent $\gamma$.  

We define the $1/e$ dephasing rates $\Gamma_{N,\lambda}$, for each dephasing channel as

\begin{equation}
\label{Gam}
\Gamma_{N,\lambda} =  \left[(2\pi)^{\gamma-1}\kappa_\lambda^2 A_\lambda^2\int_0^\infty \frac{\mathrm{d}z}{z^\gamma}g_N(z)\right]^{1/(1+\gamma)} \equiv\left[\kappa_\lambda A_\lambda \eta_N^{1/2} \right]^{2/(1+\gamma)},
\end{equation}

\noindent where the numerical factors $\eta_{0},\eta_{1}$ depend on the noise exponent $\gamma$ and the Ramsey and echo filter functions and are defined as

\begin{equation}
\label{etadef}
\eta_N =  (2\pi)^{\gamma-1}\int_0^\infty \frac{\mathrm{d}z}{z^\gamma}g_N(z).
\end{equation}

\noindent As discussed in~\cite{byla11}, for the case of $\gamma=1$, these factors are given by

\begin{align}
\eta_0 &\approx \mathrm{ln}\left(\frac{1}{\omega_\mathrm{low}t}\right)\\
\eta_1 &= \mathrm{ln}(2),
\end{align}

\noindent where $\omega_\mathrm{low}$ is the lower cutoff frequency set by the total time of all experimental iterations and $t$ is the typical free evolution time during a single experimental iteration.  Note that $\eta_1$ is completely independent of the cutoff frequency, thus avoiding any ambiguity in choosing $\omega_\mathrm{low}$ and $t$ when analyzing echo experiments, while $\eta_0$ varies only weakly with $\omega_\mathrm{low}t$ for realistic measurement settings.

\begin{figure*}
\begin{center}
\includegraphics[scale=1]{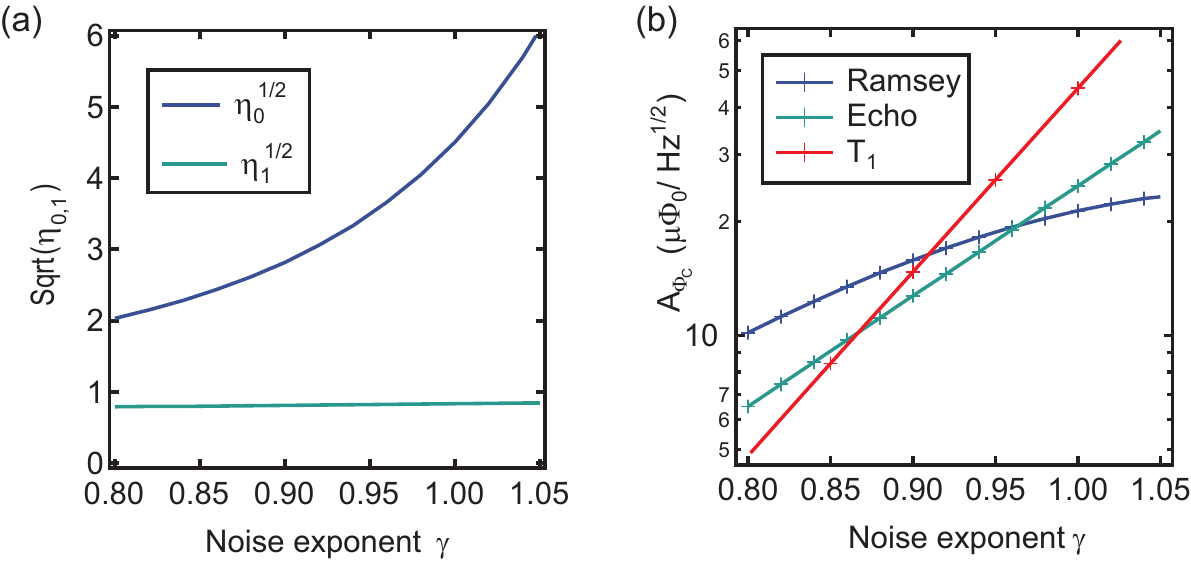}
\caption{\textbf{Flux noise analysis} (a)$\sqrt{\eta_{0,1}}$ vs $\gamma$, determined through numerical integration.  When calculating $\eta_0$, we have assumed $\omega_\mathrm{low}/2\pi = 3~\mathrm{mHz}$ and $t=200~\mathrm{ns}$. (b) Estimated coupler flux noise amplitude based measured Ramsey, Echo, and $T_1$ times, as a function of $\gamma$.}
\end{center}
\label{fig:NoiseAnaly}
\end{figure*}

For $\gamma \neq 1$, we determine the numerical factors
through numerical integration of equation~\eqref{etadef}, as discussed in reference~\cite{sank12}.  For the Ramsey sequence, 

\begin{equation}
\eta_0 = (2\pi)^{\gamma-1}\int_{\omega_\mathrm{low}t}^\infty\frac{\mathrm{d}z}{z^\gamma}\left(\frac{\mathrm{sin}(z/2)}{z/2}\right)^2
\end{equation}

\noindent and for the Echo sequence,

\begin{equation}
\eta_1 = (2\pi)^{\gamma-1}\int_{0}^\infty\frac{\mathrm{d}z}{z^\gamma}\left(\frac{\mathrm{sin}(z/4)}{z/4}\right)^2 \mathrm{sin}^2(z/4)
\end{equation}

Figure S4a shows $\sqrt{\eta_{0,1}}$ as a function of $\gamma$ for $\omega_\mathrm{low}/2\pi = 3~\mathrm{mHz}$ and $\tau = 200~\mathrm{ns}$.

\subsection{Estimating the flux noise amplitude in the coupler loop}
\label{est}

We now combine the results of the previous sections with our qubit coherence measurements to estimate the flux noise amplitude and exponent in the coupler loop.  We first consider the Ramsey and Echo results presented in Figure 4e,f. We define the total $1/e$ decay rates for Ramsey and Echo experiments as $\Gamma_\mathrm{0} \equiv 1/T_2^{\mathrm{Ramsey}}$ and $\Gamma_\mathrm{1} \equiv 1/T_2^\mathrm{Echo}$, respectively. We separate the decay rates into two contributions: $\Gamma_{N,\mathrm{\Phi_C}}$ due to flux noise in the coupler loop and $\Gamma_{N,\mathrm{other}}$, which includes the effect of $T_1$ as well as any additional dephasing.

When the coupler is biased far from degeneracy, $\Gamma_{N,\mathrm{\Phi_C}}$ is negligible, and thus $\Gamma_{N} = \Gamma_{N,\mathrm{other}}$.  For simplicity, we model $\Gamma_{N,\mathrm{other}}$ as exponential decay~\footnote{In reality, when biased away from coupler degeneracy, the qubit dephasing is somewhat non-exponential. However, it is difficult to quantify the non-exponential decay using existing data, and any non-exponential corrections to $\Gamma_{N,\mathrm{other}}$ would only have a small impact on our estimation of $A_{\Phi_\mathrm{C}}$.}. For arbitrary coupler bias, the total phase decay takes the form

\begin{equation}
\mathrm{exp}[-\Gamma_{N,\mathrm{other}}\tau-(\Gamma_{N,\mathrm{\Phi_C}}\tau)^{1+\gamma}].
\end{equation}

Thus, we can determine $\Gamma_{N,\mathrm{\Phi_C}}$ from the measured values of $\Gamma_{N}$ and $\Gamma_{N,\mathrm{other}}$ through the relation

\begin{equation}
\Gamma_{N,\mathrm{\Phi_C}} = \Gamma_N\left(1-\frac{\Gamma_{N,\mathrm{other}}}{\Gamma_N}\right)^{1/(1+\gamma)}
\end{equation}

Finally, from equation~\ref{Gam} the spectral density of flux noise in the coupler is given by 

\begin{equation}
\label{cohA}
A_{\Phi_\mathrm{C}}= \kappa_{\Phi_\mathrm{C}}^{-1}\eta_N^{-1/2}(\Gamma_{N,\Phi_\mathrm{C}})^{(1+\gamma)/2},
\end{equation}

\noindent where $\kappa_{\Phi_\mathrm{C}}\approx \kappa_{\Delta_B,\Phi_\mathrm{C}}$ is experimentally determined from the slope of $\Delta_\mathrm{B}$ vs $\Phi_\mathrm{C}$ (FIG. 4a,b).  

In Figure 10b we plot the value of $A_{\Phi_\mathrm{C}}$ that fits best to our Ramsey and Echo measurements using equation~\ref{cohA}, and to our $T_1$ measurements using equation~\ref{T1C}, for different values of $\gamma$. Although we are unable to choose values of  $A_{\Phi_\mathrm{C}}$ and $\gamma$ that fit perfectly with all three measurements, they are roughly bounded within the triangular region between the three curves in Figure 10b, where $10~\mu \Phi_0/\sqrt{\mathrm{Hz}}<A_{\Phi_\mathrm{C}}<19~\mu \Phi_0/\sqrt{\mathrm{Hz}}$ and $0.86<\gamma<0.96$.  For concreteness,  when plotting theory curves in the main text we choose $\gamma = 0.91$ and $A_{\Phi_\mathrm{C}} = 15~\mu \Phi_0/\sqrt{\mathrm{Hz}}$, which results in reasonably good agreement with all three measurements.

This estimate for the flux noise in the coupler loop is substantially larger than the value previously reported for flux qubits made with the same fabrication process, where we measured a flux noise amplitude of $1.4$ $\mu\Phi_0/\sqrt{\text{Hz}}$~\cite{yan16}.  The most significant difference between the coupler loop and the low-noise qubit loops is  the loop size; the coupler loop is $20$ times larger in area.  Therefore, these results motivate future efforts to study the dependence of flux noise on loop size beyond the scope of this work and previous efforts~\cite{lant09}.  Such measurements would help to inform architectural choices for optimizing coherence and coupling in next-generation quantum annealers.

\section{Residual coupling with coupler biased `off'}

\begin{figure}
\begin{center}
\includegraphics[scale=1]{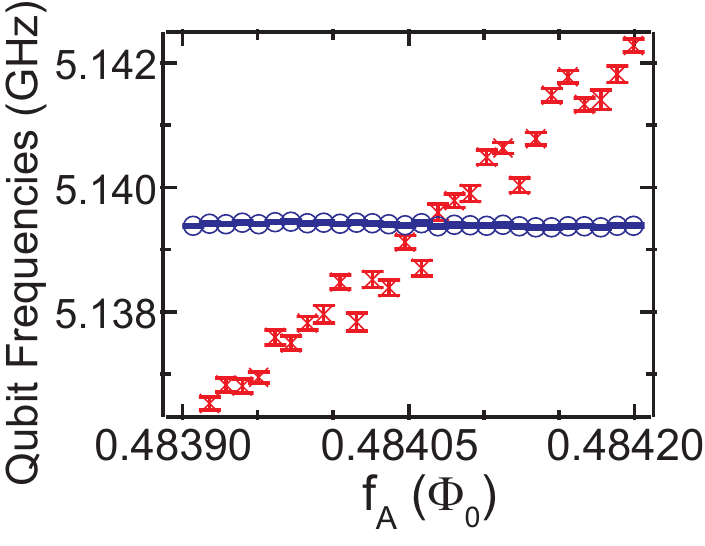}
\caption{\textbf{Zero coupling.} Detailed data for the qubit level crossing with the coupler nominally biased for zero coupling ($f_\mathrm{C} = 0.5$). Red hourglasses: Qubit A. Blue circles: Qubit B.  }
\label{fig:zero}
\end{center}
\end{figure}

Here, we describe the technique that we used to place an upper bound on any residual coupling when the coupler is nominally biased to provide zero coupling ($f_\mathrm{C} = 0.402$).  As illustrated in Figure~\ref{fig:zero}, we observe no avoided crossing in spectroscopy, allowing us to bound any nonzero residual coupling to $<220$ kHz, a limit determined by the resolution in $f_\mathrm{A}$ set by our bias current source. For each value of $f_\mathrm{A}$, the frequency of Qubit A is determined by fitting the spectroscopy trace to a Gaussian function.  Qubit B is biased on degeneracy, and its frequency is precisely determined through Ramsey spectroscopy.   

%\bibliographystyle{abbrv}
%\bibliography{ZZrefs}

\end{document}